\documentclass[a4paper,11pt]{article}
\pdfoutput=1 

\usepackage{jheppub} 

\usepackage[T1]{fontenc} 

\newcommand{\de}{\text{d}}
\newcommand{\psu}{\mathfrak{psu}}
\newcommand{\su}{\mathfrak{su}}
\newcommand{\so}{\mathfrak{so}}
\renewcommand{\u}{\mathfrak{u}}
\renewcommand{\sl}{\mathfrak{sl}}

\newcommand{\sL}{\text{L}}
\newcommand{\sR}{\text{R}}

\newcommand{\gsl}{L}
\newcommand{\gsu}{J}

\title{\boldmath S matrix for a three-parameter integrable deformation of $AdS_3\times S^3$ strings}

\author[a]{Marco Bocconcello,}
\author[b]{Isari Masuda,}
\author[b]{Fiona K.~Seibold,}
\author[a,b,c]{Alessandro Sfondrini}

\affiliation[a]{Dipartimento di Fisica e Astronomia, Universit\`a degli Studi di Padova,\\ via Marzolo 8, 35131 Padova, Italy}
\affiliation[b]{Institut f\"ur theoretische Physik, ETH Z\"urich,\\ Wolfgang-Pauli-Strasse 27, 8093 Z\"urich, Switzerland}
\affiliation[c]{Istituto Nazionale di Fisica Nucleare, Sezione di Padova,\\ via Marzolo 8, 35131 Padova, Italy}

\emailAdd{marco.bocconcello@studenti.unipd.it}
\emailAdd{masudai@student.ethz.ch}
\emailAdd{fseibold@itp.phys.ethz.ch}
\emailAdd{alessandro.sfondrini@unipd.it}

\abstract{%
We consider the three-parameter integrable deformation of the $AdS_3\times S^3$ superstring background constructed in arXiv:1811.00453. Working on the string worldsheet in uniform lightcone gauge, we find the tree-level bosonic S~matrix of the model and study some of its limits.
}

\begin{document} 
\maketitle
\flushbottom

\section{Introduction}
\label{sec:intro}

The study of exactly solvable systems has historically been very important in advancing our understanding of theoretical physics. More recently, it was understood that some string backgrounds are simple enough to be treated exactly, at least in the limit of free strings --- when one is only concerned with finding the string energy levels from the quantization of the underlying non-linear sigma model. This is not surprising for relatively simple backgrounds, such as flat-space or pp-wave geometries, but it is a very non-trivial fact for more general ones. The most famous such background is probably $AdS_5\times S^5$, which is of great interest in the context of the AdS/CFT correspondence, and can be studied exactly by integrability techniques, see \textit{e.g.}\ refs.~\cite{Arutyunov:2009ga, Beisert:2010jr,Bombardelli:2016rwb}.
Another very interesting family of integrable backgrounds are those of the $AdS_3\times S^3$ type, see ref.~\cite{Sfondrini:2014via} for a review,%
\footnote{%
Here we will be talking about $AdS_3\times S^3\times T^4$, and in fact since we are concerned with classical features of the background we will mostly ignore the $T^4$ factor. It is worth noting that much of what we say could apply also to the somewhat more involved $AdS_3\times S^3\times S^3\times S^1$ background, see refs.~\cite{Babichenko:2009dk,Sundin:2012gc} and~\cite{Borsato:2012ud,Borsato:2012ss,Borsato:2015mma,Dei:2018jyj} for studies of its classical and quantum integrability, respectively.}
which have a few special features. While the $AdS_5\times S^5$ is supported by a (self-dual) Ramond-Ramond (RR) flux, the  $AdS_3\times S^3$ backgrounds may be supported by a combination of RR and Neveu-Schwarz-Neveu-Schwarz (NSNS) fluxes, effectively yielding a one-parameter family of backgrounds with ``mixed fluxes'', all classically integrable~\cite{Cagnazzo:2012se}. Limiting cases of this family are the pure-RR background (where there are no NSNS fluxes) or the pure-NSNS ones (no RR fluxes). The former is most similar to $AdS_5\times S^5$ and was the first to be studied by integrability~\cite{Babichenko:2009dk,Sundin:2012gc,Borsato:2012ud,Beccaria:2012kb,Borsato:2013qpa,Borsato:2013hoa}, while the latter corresponds to a Wess-Zumino-Witten (WZW) model. In the WZW case, the spectrum can be worked out by using the representation theory of the underlying $\sl(2)\oplus\su(2)$ Ka\v{c}-Moody algebra~\cite{Maldacena:2000hw,Maldacena:2000kv,Maldacena:2001km} as well as by integrability~\cite{Baggio:2018gct,Dei:2018mfl}.

While the landscape of exactly solvable $AdS$ backgrounds is already quite rich --- and still in the process of being thoroughly explored --- it is very remarkable that even more general integrable models may be considered by constructing integrable deformations of the underlying non-linear sigma models.
Some early progress in constructing integrable deformations was achieved for the principal chiral model~\cite{Fateev:1996ea}, see also refs.~\cite{Lukyanov:2012zt} for more recent work on the subject.
 A very general framework to obtain such deformations is that of \textit{Yang-Baxter deformation} originally introduced by Klimcik~\cite{Klimcik:2002zj,Klimcik:2008eq}.
In general, one distinguishes between homogeneous and inhomogeneous Yang-Baxter deformations. The former have a simpler geometrical interpretations, which encompasses ``TsT'' transformations~\cite{Frolov:2005ty,Frolov:2005dj} and non-abelian T~dualities, see refs.~\cite{Matsumoto:2014nra,Matsumoto:2015jja,vanTongeren:2015soa,Osten:2016dvf,Borsato:2016pas}.
The inhomogeneous case, and in particular its application to superstring models, is the focus of this article, see refs.~\cite{Delduc:2013fga,Delduc:2013qra,Kawaguchi:2014qwa, Hoare:2014pna}.
  In the simplest case, a deformation introduces a single new parameter in the non-linear sigma model,%
\footnote{%
This is on top of the overall scale of the space, given \textit{e.g.}\ by the radius of the sphere in units of the string length, which we shall mostly keep implicit.}
though multi-parameter deformations such as the ones we will consider here are possible.
As a rule, while the integrable structure of the models arising from inhomogeneous deformations is under very good control, their geometric properties and the features of their would-be dual quantum theory --- provided that the AdS/CFT correspondence may be extended to accommodate such deformations --- are not well understood.

It turns out that the case of $AdS_3\times S^3$ is particularly rich also when it comes to deformations. As we mentioned, the undeformed model featured \textit{two} parameters: the overall size of the geometry, and the ratio between RR and NSNS fluxes.%
\footnote{%
In the case of $AdS_3\times S^3\times S^3\times S^1$ we would encounter one more parameter, which is the ratio of the radii of the two spheres.}
When it comes to applying the procedure by Klimcik to this background, it is possible to introduce from the get-go a \textit{bi-Yang-Baxter} deformation like in refs.~\cite{Klimcik:2008eq,Klimcik:2014bta}. This is because the (super)isometries of the undeformed background have a direct sum structure, $\psu(1,1|2)\oplus\psu(1,1|2)$, which is directly related to the chiral-antichiral split in the dual CFT. Therefore, the Yang-Baxter deformation may be applied independently to either copy of $\psu(1,1|2)$, as it was done in ref.~\cite{Hoare:2014oua}. Additionally, one can also accommodate the Wess-Zumino (WZ) term in the deformation~\cite{Delduc:2014uaa}, which allows to incorporate all $AdS_3\times S^3$ mixed-flux backgrounds within a \textit{three-parameter} family of deformations, as it was done in ref.~\cite{Delduc:2018xug}.
It is worth stressing that it is presently unknown whether this three-parameter deformation yields a superstring background. Indeed even for one- and two-parameter deformations only recently it was understood how to apply the Yang-Baxter procedure in such a way to generate a supergravity background, see refs.~\cite{Arutyunov:2015qva,Borsato:2016ose,Hoare:2018ngg} and~\cite{Seibold:2019dvf}, respectively.

While the action of this trice-deformed model has been constructed, it is an open problem to work out its deformed spectrum. Following the roadmap which proved successful \textit{e.g.}\ for $AdS_5\times S^5$, we expect that the best way to construct the spectrum is to consider the model in uniform lightcone gauge~\cite{Arutyunov:2004yx,Arutyunov:2005hd,Arutyunov:2006gs}. In this way, the two-dimensional integrable structure on the worldsheet is directly related to the string spectrum in target space. As it turns out, much like in the well-established two-dimensional integrable bootstrap approach for relativistic theories~\cite{Zamolodchikov:1978xm}, the best way to understand the spectrum is to avoid dealing with the Hamiltonian, and focus instead on the S~matrix on the string worldsheet. This will be constrained by the lightcone gauge symmetries as well as by integrability.
Indeed, in many cases --- see \textit{e.g.}\ refs.~\cite{Arutyunov:2006ak,Arutyunov:2006yd} for $AdS_5\times S^5$ and%
\footnote{%
The $AdS_5/CFT_4$ integrable S~matrix was first obtained by Beisert by a bootstrap approach on in $\mathcal{N}=4$ supersymmetric Yang-Mills theory~\cite{Beisert:2005tm}, rather than on the string worldsheet.}
 refs.~\cite{Borsato:2014exa,Borsato:2014hja} for $AdS_3\times S^3\times T^4$ --- this is enough to fix the S~matrix almost uniquely.
In the case of inhomogeneous Yang-Baxter deformations of $AdS_5\times S^5$ it was found~\cite{Delduc:2014kha} that the original symmetry is deformed to a \textit{quantum-group symmetry} of the type proposed by Beisert and Koroteev~\cite{Beisert:2008tw}. For the case at hand it is not immediately clear how the bootstrap procedure may be employed to fix the trice-deformed $AdS_3\times S^3$ S~matrix, also because the deformation relative to the Wess-Zumino term is \textit{a priori} different in structure from the usual Yang-Baxter ones.
In any case, it is important to verify at every step of the way that the ``bootstrapped'' S~matrix does indeed match what can be found from perturbation theory--- for the $AdS_5\times S^5$ case this was done in refs.~\cite{Arutyunov:2013ega, Arutyunov:2015qva, Seibold:2020ywq}. Once the S~matrix is known, one proceeds to obtain the spectrum by suitable Bethe Ansatz techniques~\cite{Zamolodchikov:1989cf}, whose discussion is beyond the scope of our work.

Our main goal concerns the first step in this roadmap, namely the study of the S~matrix for the  most general three-parameter deformation of ref.~\cite{Delduc:2018xug}. In particular, below we will construct its bosonic tree-level S~matrix, which will provide an important input for the comprehensive study of this background.
The article is structured as it follows. In section~\ref{sec:nlsm} we review the geometry found in ref.~\cite{Delduc:2018xug}. In section~\ref{sec:smatrix} we present the computation of the S~matrix. In section~\ref{sec:limits} we discuss some limits of the three-parameter case: the two-parameter setup of ref.~\cite{Hoare:2014oua}, whose conjectured S~matrix we validate, and some ``chiral'' limits in which the scattering simplifies. We conclude in section~\ref{sec:conclusions}, relegating the full results of our computation for the quartic Hamiltonian and for the S~matrix itself to appendices~\ref{app:quarticH} and~\ref{app:Smatrix}, respectively.
For the reader's convenience, a \textit{Wolfram Mathematica} notebook containing the tree-level S~matrix is also attached to the arXiv submission of this paper.

\section{The non-linear sigma model}
\label{sec:nlsm}

Let us collect here the definition of the background metric and Kalb-Ramond field which we will need in what follows.

\subsection{Isometric coordinates}
From the construction or ref.~\cite{Delduc:2018xug} one can find the bosonic action by putting the Fermions to zero. This is given by the line element and $B$-field which take the form
\begin{equation}
ds^2=ds^2_{(1)}+ds^2_{(2)}\,,\qquad B=B_{(1)}+B_{(2)}\,,
\end{equation}
where the subscript labels $(1)$ and $(2)$ refer to $AdS_3$ and $S^3$, respectively, and
\begin{equation}
\label{eq:defds}
\begin{aligned}
ds^2_{(1)}=\frac{1}{F_{(1)}}\Big[\,\frac{1-q^2 \rho^2(1+\rho^2)}{1+\rho^2}\,\de \rho^2-2q \chi_{-}\rho(1+\rho^2)\, \de \rho\, \de t+2q \chi_+ \rho^3\, \de \rho\, \de \psi\qquad\quad\\
-\big(1+\chi_{-}^2(1+\rho^2)\big)(1+\rho^2)\,\de t^2+2 \chi_+\chi_- \rho^2(1+\rho^2)\,\de t\,\de\psi +\rho^2 (1 - \rho^2 \chi_+^2)\, \de \psi^2 \,\Big],\\
ds^2_{(2)}=\frac{1}{F_{(2)}}\Big[\,\frac{1+q^2 r^2(1-r^2)}{1-r^2}\,\de r^2-2q \chi_{-}r(1-r^2)\, \de r\, \de \omega-2q \chi_+ r^3\, \de r\, \de \phi \qquad\quad\\
\qquad +\big(1+\chi_{-}^2(1-r^2)\big)(1-r^2)\,\de \omega^2+2 \chi_+\chi_- r^2(1-r^2)\,\de\omega\,\de\phi+ r^2  (1 + \chi_+^2 r^2) \,\de \phi^2\,\Big],
\end{aligned}
\end{equation}
and%
\footnote{%
Our definition of the Kalb-Ramond field differs from the one given in Appendix~C of ref.~\cite{Delduc:2018xug} by a term $B_{\text{diff}}$ such that $\de B_{\text{diff}}=0$. We have taken our $B$-field to vanish when $r\to0$ and $\rho\to0$, which is convenient for what follows.}
\begin{equation}
\label{eq:defB}
\begin{aligned}
&&B_{(1)}=\frac{a\,q}{F_{(1)}}\,\rho^2\Big[2+(1+\rho^2)q^2+(1+\rho^2)\chi_-^2+(1-\rho^2)\chi_{+}\Big]\,
\de t\wedge \de \psi\,,\\
&&B_{(2)}=\frac{a\,q}{F_{(2)}}\,r^2\Big[2+(1-r^2)q^2+(1-r^2)\chi_-^2+(1+r^2)\chi_{+}\Big]\,
\de \omega\wedge \de \phi\,,
\end{aligned}
\end{equation}
with
\begin{equation}
\label{eq:defA}
\begin{aligned}
F_{(1)}=&\,1-\chi_{+}^2\rho^2+\chi_-^2(1+\rho^2)-q^2\rho^2(1+\rho^2)\,,\\
F_{(2)}=&\,1+\chi_{+}^2r^2+\chi_-^2(1-r^2)+q^2r^2(1-r^2)\,,
\end{aligned}
\end{equation}
and
\begin{equation}
\label{eq:defa}
a=\frac{1}{\sqrt{\big(q^2+\chi_+^2+\chi_-^2\big)^2+4\big(q^2-\chi_+^2\chi_-^2\big)}}\,.
\end{equation}
The dilaton and RR fluxes were not given in ref.~\cite{Delduc:2018xug}. Two possible solutions for the dilaton, which extend those found in ref.~\cite{Seibold:2019dvf} are, up to a constant dilaton~$\Phi_0$, 
\begin{equation}
\label{eq:dilaton}
\begin{aligned}
&e^{2\Phi}=e^{2\Phi_0}\,\frac{P(\rho,r)^2}{F_{(1)}\,F_{(2)}}\,,\\
\text{with}\quad&P(\rho,r)=1+\chi_{-}^2-\rho^2\,r^2\,(\chi_+^2-\chi_-^2)\,,\\
\text{or}\quad&P(\rho,r)=1+\chi_{+}^2-(1+\rho^2)(1-r^2)(\chi_+^2-\chi_-^2)\,.
\end{aligned}
\end{equation}
It is worth emphasising that, despite its rather complicated form, the background has four shift isometries relative to the coordinates $t, \psi, \omega$ and $\phi$, under \textit{e.g.}\ $t\to t+\text{const.}$, and so~on. Moreover, just like in the undeformed case, the $AdS_3$ and $S^3$ parts of the trice-deformed metric and $B$ field are related by analytic continuation,
\begin{equation}
\label{eq:analyticcont}
r\leftrightarrow-i \rho\,,\qquad \omega\leftrightarrow t\,,\qquad
\phi\leftrightarrow \psi\,,\qquad 
B_{(1)}\leftrightarrow-B_{(2)}\,,\qquad
\de s^2_{(1)}\leftrightarrow-\de s^2_{(2)}\,,
\end{equation}
while the two solutions for~$P(\rho,r)$ go to each other in eq.~\eqref{eq:dilaton}.

As it can be seen from these formulae, the background depends on the parameters $\chi_{+},\chi_{-}$ and $q$. The dependence on the radius of $AdS_3$ and $S^3$ is absorbed in the definition of the coordinates and will not be indicated. Let us briefly comment on the interpretation of the  three deformation parameters. Recall that the super-isometries of $AdS_3\times S^3$ factorise as $\psu(1,1|2)_{\sL}\oplus\psu(1,1|2)_{\sR}$, where ``L'' and ``R'' stand for ``left'' and ``right'' in the dual CFT${}_{2}$. It is possible to construct an integrable deformation of the resulting coset geometry with respect to either copy of the algebra~\cite{Hoare:2014oua}. The parameters $\chi_{\pm}$ are related to such a deformation, as it is simplest to see when $q=0$. Then, when $\chi_-=0$ and $\chi_+\neq0$, the two copies of $\psu(1,1|2)$ are deformed in a \textit{symmetric} way, whereas when $\chi_-\neq 0$ and $\chi_+=0$ they are deformed in an \textit{antisymmetric} way. In fact, writing $\chi_{\pm}=\tfrac{1}{2}(\chi_{\sL}\pm\chi_{\sR})$ we have that $\chi_{\sL}$ deforms $\psu(1,1|2)_{\sL}$ and $\chi_{\sR}$ deforms $\psu(1,1|2)_{\sR}$ .
The third parameter~$q$ accommodates the possibility of modifying the action by adding a Wess-Zumino term. Indeed, sending $\chi_{\pm}\to0$ while tuning~$q\to0$
one may recover the action of the $\sl(2)\oplus\su(2)$ Wess-Zumino-Witten model, or more generally the action of the ``mixed flux'' $AdS_3\times S^3$ background. To this end, we want to send $\chi_\pm\to0$ and $q\to0$ in such a way that $a q$ has a finite, non-zero limit --- specifically, $aq\to \tilde{q}/2$, where $0<\tilde{q}\leq 1$ measures the amount of NSNS fluxes relative to the RR ones~\cite{Delduc:2018xug}.

\subsection{Stereographic coordinates}
For later convenience, let us consider a new set of coordinates by setting
\begin{equation}
\begin{aligned}
\rho=&\,\sqrt{(X^1)^2+(X^2)^2}\,,\qquad &\psi = &-\text{atan}\Big(\frac{X^2}{X^1}\Big)\,,\\
r=&\,\sqrt{(X^3)^2+(X^4)^2}\,,\qquad &\phi = &+\text{atan}\Big(\frac{X^4}{X^3}\Big)\,.
\end{aligned}
\end{equation}
The advantage of this choice of coordinates is that $X^1$ and $X^2$ can be more readily used to construct charge eigenstates under the $\u(1)$ symmetry generated by $\psi$ (and similarly for $X^3,X^4$ and $\phi$) and as such they will be easier to relate to the fundamental excitations scattered by the S~matrix.

\section{The S~matrix from the lightcone gauge Hamiltonian}
\label{sec:smatrix}
The computation of the tree-level bosonic S~matrix can be done in a relatively straightforward way. We will carry it out in the first-order formalism in lightcone gauge. The first step is to work out the lightcone Hamiltonian, which we will do in this section.
For further details we refer the reader to the review~\cite{Arutyunov:2009ga} as well as to ref.~\cite{Lloyd:2014bsa} where the case of mixed-flux $AdS_3\times S^3$ is worked out.

\subsection{First-order action}
We begin by considering the non-linear sigma-model action%
\footnote{%
We omit the overall dependence on the string tension.}
\begin{equation}
S=-\frac{1}{2}\int\limits_{-\infty}^{+\infty}\de \tau\int\limits_{-R}^{+R} \de\sigma\,\Big(\gamma^{ab}\partial_a X^\mu\partial_b X^\nu G_{\mu\nu}(X)+\epsilon^{ab}\partial_a X^\mu\partial_b X^\nu B_{\mu\nu}(X)\Big)
\end{equation}
where $\gamma^{ab}=\sqrt{-h}h^{ab}$ is the unit-determinant worldsheet metric, $\epsilon^{ab}$ is the Levi-Civita tensor, and we introduced $X^\mu=(t,X^1,X^2,X^3,X^4,\omega)$. This can be recast in first-order form by introducing the conjugate momenta
\begin{equation}
P_{\mu} = \frac{\delta\, S}{\delta\, \partial_\tau X^{\mu} }\,,
\end{equation}
taking the form
\begin{equation}
\label{eq:1storderaction}
\begin{aligned}
S=&\,\int\limits_{-\infty}^{+\infty}\de \tau\int\limits_{-R}^{+R} \de\sigma
\Big(P_\mu \dot{X}^\mu +\frac{\gamma^{01}}{\gamma^{00}} C_1+\frac{1}{2\gamma^{00}}C_2\Big)\,,\\
C_1 =&\, P_\mu \acute{X}^\mu\,,\\
C_2 =&\, G^{\mu\nu}P_\mu P_\nu+G_{\mu\nu} \acute{X}^\mu\acute{X}^\nu+2G^{\mu\nu}B_{\nu\kappa}P_\mu \acute{X}^\kappa+G^{\mu\nu}B_{\mu\kappa}B_{\nu\lambda}\acute{X}^\kappa\acute{X}^\lambda\,,\\
\end{aligned}
\end{equation}
where we have highlighted the Virasoro constraints $C_1$ and $C_2$, introduced the short-hand notation $\dot{X}^\mu=\partial_\tau X^\mu$ and $\acute{X}^\mu=\partial_\sigma X^\mu$, and omitted the $X$-dependence of the metric and of the $B$-field.

\subsection{Uniform lightcone gauge fixing}
Let us introduce lightcone coordinates $X^\pm$, as well as their conjugate momenta $P_{\pm}$, by setting
\begin{equation}
\begin{aligned}
t=&\,X^+ -\alpha\, X^-\,,\qquad &\omega=&\,X^+ + (1-\alpha)\, X^-\,,\\
P_t=&\,(1-\alpha)\,P_+ -P_{-}\,,\qquad
&P_\omega =&\, \alpha\,P_+ +P_{-}\,,
\end{aligned}
\end{equation}
where $\alpha$ is a real parameter whose significance we will discuss below, around eq.~\eqref{eq:length}.
We can fix the uniform lightcone gauge~\cite{Arutyunov:2004yx,Arutyunov:2005hd,Arutyunov:2006gs} by setting
\begin{equation}
\label{eq:gaugefixing}
X^+ = \tau\,,\qquad P_- = 1\,.
\end{equation}
In this way, discarding the total $\tau$-derivative $\int\de\tau \dot{X}^-  $, the action~\eqref{eq:1storderaction} becomes simply
\begin{equation}
\label{eq:gaugefixedaction}
S=\int\limits_{-\infty}^{+\infty}\de \tau\int\limits_{-R}^{+R} \de\sigma\Big[ P_j \dot{X}^j+ P_+(X^j,\acute{X}^j,P_j)\Big]\,,
\end{equation}
depending only on the transverse fields $X^j$, $j=1,\dots 4$, provided that the remaining longitudinal fields satisfy the Virasoro constraints
\begin{equation}
C_1 =0\,,\qquad C_2 = 0\,.
\end{equation}
In particular, $C_1=0$ can be solved easily by setting
\begin{equation}
\acute{X}^- = - P_j\acute{X}^j\,,
\end{equation}
while $C_2=0$ gives a quadratic equation for $P_+$. Furthermore, from the form of the action~\eqref{eq:gaugefixedaction} it follows that the lightcone Hamiltonian is
\begin{equation}
\label{eq:lcH}
H(X^j,\acute{X}^j,P_j)= - \int\limits_{-R}^{R}\de\sigma P_+(X^j,\acute{X}^j,P_j)\,.
\end{equation}

It is also worth expressing the lightcone charges in function of the original momenta $P_t$ and $P_\omega$. We have that
\begin{equation}
H= -\int\limits_{-R}^{+R}\de\sigma P_+ = \int\limits_{-R}^{+R}\de\sigma (-P_t-P_\omega)\,.
\end{equation}
Recall that the isometries of $AdS_3\times S^3$ are $\so(2,2)\oplus\so(4)$, and that they decompose into a direct sum~$\sl(2)_{\sL}\oplus\sl(2)_{\sR}\oplus \su(2)_{\sL}\oplus\su(2)_{\sR}$. Naming the Cartan elements of these four algebras as $\gsl_{\sL}$, $\gsl_{\sR}$, $\gsu_{\sL}$ and $\gsu_{\sR}$ respectively (we use $\gsl$s for the $\sl(2)$ generators and $\gsu$s for $\su(2)$ ones), it follows from the above decomposition that $-\int\de\sigma P_t= \gsl_{\sL}+\gsl_{\sR}$ and $\int\de\sigma P_\omega= \gsu_{\sL}+\gsu_{\sR}$, so that
\begin{equation}
\label{eq:Hcharges}
H= (\gsl_{\sL}+ \gsl_{\sR}) - (\gsu_{\sL}+ \gsu_{\sR})\,.
\end{equation}
The BPS bound of $\psu(1,1|2)$ is $\gsl\geq \gsu$, so that for $\psu(1,1|2)_{\sL}\oplus\psu(1,1|2)_{\sR}$ eq.~\eqref{eq:Hcharges} ensures that $H>0$ on all states except for those in short representations, for which $H=0$. As for $P_-$, from the gauge-fixing condition~\eqref{eq:gaugefixing} that
\begin{equation}
\int\limits_{-R}^{+R}\de\sigma\,P_{-}=2R\,.
\end{equation}
On the other hand,
\begin{equation}
\label{eq:length}
\int\limits_{-R}^{+R}\de\sigma\,P_{-}=\int\limits_{-R}^{+R}\de\sigma\Big[-\alpha\,P_t+(1-\alpha)\,P_\omega\Big]=\gsu_{\sL}+\gsu_{\sR}+\alpha\,H\,.
\end{equation}
Therefore, after gauge fixing, the worldsheet size is fixed in terms of the charges of the state. In the case~$\alpha=0$, the length is quantised and we can think of different choices of~$R$ as of different superselection sectors. For generic $\alpha$, instead, the length depends on the energy of a given state. This is reminiscent of $T\bar{T}$ deformations, as noted in ref.~\cite{Baggio:2018gct} and further detailed in refs.~\cite{Frolov:2019nrr,Frolov:2019xzi,Sfondrini:2019smd}. In what follows, we shall keep $\alpha$ general. As it turns out, the choice $\alpha=1/2$ yields slightly simpler formulae in most cases.

\subsection{Perturbative expansion of the Hamiltonian}
It is now straightforward to compute $H$ from~\eqref{eq:lcH} by solving $C_2=0$, \textit{cf.}~\eqref{eq:1storderaction}. It is convenient to express $H$ as an expansion in the transverse fields~$X^j,\acute{X}^j$ and $P_j$,
\begin{equation}
\label{eq:expansion}
H = H^{(2)} + H^{(4)} + \dots\,,
\end{equation}
where in the case at hand there are no odd-order terms in the expansion. Eq.~\eqref{eq:expansion} can be related to a large-tension expansion in string theory, see \textit{e.g.}~\cite{Arutyunov:2009ga}. The quadratic part $H^{(2)}$ will correspond to the free theory which emerges in the Berenstein-Maldacena-Nastase (BMN) limit~\cite{Berenstein:2002jq}.

Before proceeding with the expansion of the Hamiltonian, in our case it is convenient to first perform the following canonical transformation:
\begin{equation}
\label{eq:canonical}
P_j = \frac{\widetilde{P}_j}{\sqrt{1+\chi_-^2}}-\frac{q\,\chi_-}{\sqrt{1+\chi_-^2}} \widetilde{X}^j\,,\qquad X^j=\sqrt{1+\chi_-^2}\,\widetilde{X}^j\,.
\end{equation}
This has the effect of getting rid of what would be a total $\tau$-derivative in the Lagrangian formalism, of the form $\partial_\tau (X^jX^j)$, as well as of canonically normalising the free action and Hamiltonian.
It is also convenient to introduce complex fields as it follows:
\begin{equation}
\begin{aligned}
&\widetilde{X}^1=\frac{Z-\bar{Z}}{i\sqrt{2}}\,,\qquad
&&\widetilde{X}^2=\frac{Z+\bar{Z}}{-\sqrt{2}}\,,\qquad
&&\widetilde{X}^3=\frac{Y+\bar{Y}}{-\sqrt{2}}\,,\qquad
&&\widetilde{X}^4=\frac{Y-\bar{Y}}{i\sqrt{2}}\,,\\
&\widetilde{P}_1=\frac{P_z-\bar{P}_z}{i\sqrt{2}}\,,\quad
&&\widetilde{P}_2=\frac{P_z-+\bar{P}_z}{-\sqrt{2}}\,,\quad
&&\widetilde{P}_3=\frac{P_y+\bar{P}_y}{-\sqrt{2}}\,,
&&\widetilde{P}_4=\frac{P_y-\bar{P}_y}{i\sqrt{2}}\,.
\end{aligned}
\end{equation}
Then, the quadratic Hamiltonian reads
\begin{equation}
\label{eq:quadraticH}
\begin{aligned}
H^{(2)}=&\, P_z \bar{P}_z+\acute{Z}\acute{\bar{Z}}+m^2 Z\bar{Z}+ i \chi_+\chi_- (Z\bar{P}_z-\bar{Z}P_z)- i\lambda(Z\acute{\bar{Z}}-\bar{Z}\acute{Z})\\
&+P_y \bar{P}_y+\acute{Y}\acute{\bar{Y}}+m^2 Y\bar{Y}+ i \chi_+\chi_- (Y\bar{P}_y-\bar{Y}P_y)- i\lambda(Y\acute{\bar{Y}}-\bar{Y}\acute{Y})\,,
\end{aligned}
\end{equation}
with
\begin{equation}
\label{eq:defmlambda}
m^2=q^2+(1+\chi_+^2)(1+\chi_-^2)\,,\qquad
\lambda=a\,q\,(2+q^2+\chi_-^2+\chi_+^2)\,.
\end{equation}
It is interesting to note that, with respect to the standard action of two complex massive bosons, here we have two modifications. Firstly, there is a parity-breaking term, related to the Wess-Zumino term, which like the $B$-field itself is proportional to $aq$, \textit{cf.}\ eq.~\eqref{eq:defA}. Next, we have a time-reversal-breaking term due to the bi-Yang-Baxter deformation and proportional to $\chi_+\chi_-$.

In a similar way it is possible to work out the quartic-order Hamiltonian~$H^{(4)}$. Its expression is somewhat bulky and we collect it in appendix~\ref{app:quarticH}.

\subsection{Creation and annihilation operators}
In order to diagonalise the quadratic Hamiltonian, it is convenient to introduce oscillators $a^\dagger_p$ and $a_p$ which in the quantum theory will be promoted to creation and annihilation operators. The procedure follows the standard construction for a free complex scalar field, with some minor modifications due to the parity and time-reversal violating terms in eq.~\eqref{eq:quadraticH}. We start by observing that the free wave equation related to~\eqref{eq:quadraticH} can be solved by the ansatz
\begin{equation}
\label{eq:oscillators}
\begin{aligned}
Z(\sigma) = \int\de^2\sigma\Bigg[ \frac{e^{-i \omega(p) \tau +i p\sigma}}{\sqrt{2g(p)}}a^z(p)+\frac{e^{+i \bar{\omega}(p) \tau -i p\sigma}}{\sqrt{2\bar{g}(p)}}{a}_{\bar{z}}^\dagger(p)\Bigg]\,,\\
\bar{Z}(\sigma) = \int\de^2\sigma\Bigg[ \frac{e^{-i \bar{\omega}(p) \tau +i p\sigma}}{\sqrt{2\bar{g}(p)}}{a}^{\bar{z}}(p)+\frac{e^{+i {\omega}(p) \tau -i p\sigma}}{\sqrt{2{g}(p)}}a_z^\dagger(p)\Bigg]\,,\\
Y(\sigma) = \int\de^2\sigma\Bigg[ \frac{e^{-i \omega(p) \tau +i p\sigma}}{\sqrt{2g(p)}}a^y(p)+\frac{e^{+i \bar{\omega}(p) \tau -i p\sigma}}{\sqrt{2\bar{g}(p)}}{a}_{\bar{y}}^\dagger(p)\Bigg]\,,\\
\bar{Y}(\sigma) = \int\de^2\sigma\Bigg[ \frac{e^{-i \bar{\omega}(p) \tau +i p\sigma}}{\sqrt{2\bar{g}(p)}}{a}^{\bar{y}}(p)+\frac{e^{+i {\omega}(p) \tau -i p\sigma}}{\sqrt{2{g}(p)}}a_y^\dagger(p)\Bigg]\,.
\end{aligned}
\end{equation}
Using the equations of motion, we find that it must be
\begin{equation}
\label{eq:omega}
\omega(p)=\sqrt{p^2-2\lambda p +m^2}-\chi_+\chi_-\,,\qquad
\bar{\omega}(p)=\sqrt{p^2+2\lambda p +m^2}+\chi_+\chi_-\,,
\end{equation}
where $m$ and $\lambda$ are given by eq.~\eqref{eq:defmlambda}.
The expressions for $P_z, \bar{P}_z, P_y$ and $\bar{P}_y$ follow from Hamilton's equations. Finally, requiring the fields and their momenta to be canonically conjugate. as well as the oscillators to satisfy canonical relations, we find the normalisation to be
\begin{equation}
g(p)= \omega(p)+\chi_+\chi_-\,,\qquad
\bar{g}(p)= \bar{\omega}(p)-\chi_+\chi_-\,.
\end{equation}
The quadratic Hamiltonian then has the standard form
\begin{equation}
H^{(2)} = \int\de p \Big[\omega(p)\Big(a_z^\dagger(p) a^z(p)+a_y^\dagger(p) a^y(p)\Big)
+
\bar{\omega}(p)\Big({a}_{\bar{z}}^\dagger(p) {a}^{\bar{z}}(p)+{a}_{\bar{y}}^\dagger(p) {a}^{\bar{y}}(p)\Big)\Big]\,.
\end{equation}
Moreover, the particles are distinguished by the $\u(1)$ charges $J_{\psi}$ and $J_{\phi}$, relative to the shift isometries in $\psi$ and $\phi$, \textit{cf.}~\eqref{eq:defds}. These are the spin in $AdS_3$ and $S^3$, respectively. We list the charges in table~\ref{tab:charges}.

\begin{table}[tbp]
\centering
\begin{tabular}{|ccccc|}
\hline
Oscillator& Particle & $J_\psi$ & $J_\phi$ & $H$\\
\hline 
$a^\dagger_z(p)$ & $|z(p)\rangle$ &  $-1$ & $0$ & $\omega(p)$\\
${a}^\dagger_{\bar{z}}(p)$ & $|\bar{z}(p)\rangle$ &  $+1$ & $0$ & $\bar{\omega}(p)$\\
$a^\dagger_y(p)$ & $|y(p)\rangle$ &  $0$ & $+1$ & $\omega(p)$\\
${a}^\dagger_{\bar{y}}(p)$ & $|\bar{y}(p)\rangle$ &  $0$ & $-1$ & $\bar{\omega}(p)$\\
\hline
\end{tabular}
\caption{
\label{tab:charges} For each of the oscillators $a^\dagger_z,a^\dagger_y, {a}_{\bar{z}}^\dagger$ and ${a}^\dagger_{\bar{y}}$, we list the particles that it creates in the quantum theory (\textit{e.g.}, $|z(p)\rangle = a^{\dagger}_z(p)|0\rangle$) as well as the particle's $\u(1)$ charges. Here $J_\psi$ is a compact Cartan element of $\so(2,2)$ which, in terms of the $\sl(2)_{\sL}\oplus \sl(2)_{\sR}$ decomposition is $J_\psi = \gsl_{\sL}- \gsl_{\sR}$, while for $\so(4)$ we have $J_\phi=\gsu_{\sL}-\gsu_{\sR}$.}
\end{table}

The computation of the tree-level S~matrix can be done in the interaction picture, and it essentially boils down to rewriting the quartic Hamiltonian in terms of oscillators, see \textit{e.g.}\ ref.~\cite{Arutyunov:2009ga}. Plugging \eqref{eq:oscillators} in the quartic Hamiltonian of appendix~\ref{app:quarticH}, we find several integrals in six variables: $\de^2\sigma$ and $\de p_1\de p_2\de p_3 \de p_4$. The first two integrals yield the energy- and momentum-conservation $\delta$-functions. The two $\delta$-functions have support only on
\begin{equation}
p_3 = p_1\quad\text{and}\quad p_4=p_2\,,\qquad
\text{or}\qquad
p_4 = p_1\quad\text{and}\quad p_3=p_2\,.
\end{equation}
Moreover, they yield a Jacobian
\begin{equation}
\Omega_{12}= \frac{1}{\omega'(p_2)-\omega'(p_1)}\,.
\end{equation}
Eventually, we are left with a sum of expressions of the form
\begin{equation}
H^{(4)} = \cdots+ \int\de p_1\de p_2\, T_{ij}^{kl}(p_1,p_2)\, a^{\dagger}_{k}(p_1)a^{\dagger}_{l}(p_2)\,a^{i}(p_1)a^{j}(p_2)+\dots\,,
\end{equation}
where~$T_{ij}^{kl}(p_1,p_2)$ is the tree-level S~matrix element and $i,j,k,l$ can be any of the flavours $z,y,\bar{z},\bar{y}$, for a total of in principle $4^4=256$ possible processes. However, it follows immediately from the conservation of the charges $J_\psi$ and $J_\phi$ that most of these processes are straightforwardly forbidden (see table~\ref{tab:charges}). It actually turns out that the tree-level S~matrix is \emph{diagonal}, \textit{i.e.}\ it takes the form
\begin{equation}
\label{eq:Tisdiagonal}
T_{ij}^{kl}(p_1,p_2)= \delta_{i}^{k}\,\delta_{j}^{l}\, T_{ij}(p_1,p_2)\,.
\end{equation}
This was not a foregone conclusion since --- based on the $\u(1)$ symmetries alone --- we could have expected \textit{e.g.}\ $T_{z\bar{z}}^{y\bar{y}}\neq0$, which is not the case. This is actually quite significant, as we will discuss below, in the paragraph around eq.~\eqref{eq:largehsmatrix}.
It should be stressed that we do not expect the S~matrix to remain diagonal beyond tree level --- this is not the case even for the simplest undeformed $AdS_3\times S^3$ backgrounds~\cite{Borsato:2012ud,Sundin:2016gqe}. Moreover, let us remark that an immediate consequence of~\eqref{eq:Tisdiagonal} is that the tree-level S matrix satisfies the classical Yang-Baxter equation as expected. We are therefore left with 16 diagonal processes $T_{ij}(p_1,p_2)$, which we have collected in appendix~\ref{app:Smatrix}.

\section{Special limits}
\label{sec:limits}
The form of the three-parameter S~matrix of appendix~\ref{app:Smatrix} is not particularly transparent. In what follows, we will restrict to some particular limits in which its structure simplifies considerably.

\subsection{Mixed-flux background}
A simple check of our construction is that we should be able to retrieve the mixed-flux $AdS_3\times S^3$ S~matrix~\cite{Hoare:2013pma,Hoare:2013lja,Lloyd:2014bsa}. In particular, we want to compare with the tree-level result of ref.~\cite{Hoare:2013pma}. By taking the limit
\begin{equation}
\chi_{\pm}\to0\,,\quad q\to0,\qquad \text{so that}\quad
q\,a\to\frac{\tilde{q}}{2}\,,
\end{equation}
we obtain the dispersion
\begin{equation}
\omega(p)=\sqrt{p^2-2\tilde{q}\,p+1}\,,\qquad
\bar{\omega}(p)=\sqrt{p^2+2\tilde{q}\,p+1}\,.
\end{equation}
The tree-level S~matrix matches perfectly with that of ref.~\cite{Hoare:2013pma}. Without reporting it all, let us give an example of a tree-level S-matrix element,
\begin{equation}
T_{zz}=(\alpha-\tfrac{1}{2})\big(p_2\,\omega(p_1)-p_1\,\omega(p_2)\big)-
\frac{p_2\,\omega(p_1)+p_1\,\omega(p_2)}{2}\frac{p_1+p_2}{p_1-p_2}\,.
\end{equation}
It is interesting to see what happens in the limit
\begin{equation}
\tilde{q}\to1\,,
\end{equation}
which corresponds to the WZW model. This S~matrix was found at tree-level in ref.~\cite{Hoare:2013pma} and further studied in refs.~\cite{Baggio:2018gct,Dei:2018mfl}. Then the dispersion relation is chiral
\begin{equation}
\omega(p)=\big|p-1\big|\,,\qquad
\bar{\omega}(p)=\big|p+1\big|\,.
\end{equation}
In order to obtain a well-defined perturbative scattering matrix, we need the two wave packets to have different group velocities $\partial \omega/\partial p$ (or $\partial\bar{\omega}/\partial p$). Let us choose $p_1$ and $p_2$ so that
\begin{equation}
\frac{\partial \omega}{\partial p}\Big|_{p_1}=+1\,,\qquad
\frac{\partial \omega}{\partial p}\Big|_{p_2}=-1\,.
\end{equation}
In that case we find
\begin{equation}
T_{zz}=(\alpha-\tfrac{1}{2})\Big(2p_1p_2-\frac{1}{2}(p_1+p_2)\Big) -\frac{1}{2}(p_1+p_2)\,.
\end{equation}
In fact, the whole S~matrix drastically simplifies and in the $\alpha=1/2$ gauge takes the form~\cite{Baggio:2018gct}
\begin{equation}
\label{eq:WZWSmat}
T_{ij}(p_1, p_2) = \frac{1}{2}\Big[
p_1 (J_\psi^{j}+J_\phi^{j})+
p_2 (J_\psi^{i}+J_\phi^{i})\Big]\,.
\end{equation}
Here $J_{\psi}^{i,j}$ and $J_{\phi}^{i,j}$ are the eigenvalues of the $\u(1)$ charges of table~\ref{tab:charges} relative to the particle flavour $i$ or $j$. This linear structure is instrumental in reproducing, from the point of view of the factorised scattering and (mirror thermodynamic) Bethe Ansatz, the WZW spectrum~\cite{Dei:2018mfl}. In this sense, it would be interesting to find other models with similar features, see section~\ref{sec:chiral}.

\subsection{Two-parameter limit}
A natural restriction is to set $q=0$, recovering the two-parameter model studied in ref.~\cite{Hoare:2014oua}. In that limit, the S~matrix had not been computed but it had been conjectured based on symmetry arguments.
To begin with, let us briefly recall how the S~matrix for the \textit{undeformed} $AdS_3\times S^3$ background can be constructed out of symmetry consideration. For further details we refer the reader to refs.~\cite{Borsato:2014exa,Borsato:2014hja} for the somewhat simpler case of Ramond-Ramond (RR) backgrounds and to ref.~\cite{Hoare:2013pma, Hoare:2013lja,Lloyd:2014bsa} for generic backgrounds (with RR and NSNS background fluxes), as well as to~\cite{Sfondrini:2014via} for a review. The supersymmetries of the S~matrix are given by half of $\psu(1,1|2)_{\sL}\oplus\psu(1,1|2)_{\sR}$, and namely by the subalgebra
\begin{equation}
\label{eq:bigalgebra1}
\{Q_{\sL}{}^i, S_{\sL}{}_{j} \} = \delta_{j}^{i}\, (\gsl_{\sL}-\gsu_{\sL}),\qquad
\{Q_{\sR}{}_{i}, S_{\sR}{}^{j} \} = \delta_{i}^{j}\, (\gsl_{\sR}-\gsu_{\sR})\,,
\end{equation}
with $i,j=1,2$, supplemented by the central extension $\gsu_{\sR}$ introduced above.
\begin{equation}
\label{eq:bigalgebra2}
\{Q_{\sL}{}^i, Q_{\sR}{}_{j} \} = \delta_{j}^{i}\, P\,,
\qquad
\{S_{\sL}{}_i, S_{\sR}{}^{j} \} = \delta^{j}_{i}\, K\,.
\end{equation}
While this central extension spoils the direct-sum form of the  $\psu(1,1|2)_{\sL}\oplus\psu(1,1|2)_{\sR}$ algebra, it is well understood that this can emerge in lightcone gauge --- see ref.~\cite{Arutyunov:2006ak,Arutyunov:2006yd} for a detailed discussion of the $AdS_5\times S^5$ case and~\cite{Borsato:2014hja} for $AdS_3\times S^4\times T^4$. For future convenience, we introduce the notation
\begin{equation}
H_{\sL}\equiv \gsl_{\sL}-\gsu_{\sL}\,,\qquad
H_{\sR}\equiv \gsl_{\sR}-\gsu_{\sR}\,,\qquad
M \equiv H_{\sL}-H_{\sR}\,,\quad
H = H_{\sL}+H_{\sR}\,.
\end{equation}
Note that $M=J_\phi-J_\psi$ is quantised and $H$ is the lightcone Hamiltonian of eq.~\eqref{eq:lcH}.

It also turns out that, for the purpose of studying the S~matrix, it is sufficient to study ``short'' (atypical representation of this algebra). For one-particle states, these consist of two Bosons and two Fermions, and on them the following shortening condition holds true:
\begin{equation}
\label{eq:shortening}
H_{\sL}\,H_{\sR}-P\,K=0\,,\qquad i.e.\quad
H^2=M^2+4\,P\,K\,.
\end{equation}
Note that this can be thought of as a dispersion relation. As a final simplification, it turns out that all the representation of interest can be obtained from the smaller algebra
\begin{equation}
\label{eq:smallalgebra}
\{Q_{\sL}, S_{\sL} \} = H_{\sL},\quad
\{Q_{\sR}, S_{\sR} \} = H_{\sR}\,,\qquad
\{Q_{\sL}, Q_{\sR} \} = P,\quad
\{S_{\sL}, S_{\sR} \} = K\,,
\end{equation}
subject to the same condition~\eqref{eq:shortening}, by setting
\begin{equation}
\label{eq:algebrasplit}
\begin{aligned}
&Q_{\sL}{}^1=Q_{\sL}\otimes 1\,,\quad
&&Q_{\sL}{}^2=1\otimes Q_{\sL}\,,\qquad
&&S_{\sL}{}_1=S_{\sL}\otimes 1\,,\quad
&&S_{\sL}{}_2=1\otimes S_{\sL}\,,\\
&Q_{\sR}{}_1=Q_{\sR}\otimes 1\,,\quad
&&Q_{\sR}{}_2=1\otimes Q_{\sR}\,,\qquad
&&S_{\sR}{}^1=S_{\sR}\otimes 1\,,\quad
&&S_{\sR}{}^2=1\otimes S_{\sR}\,.
\end{aligned}
\end{equation}
 The short representations of the algebra~\eqref{eq:smallalgebra} consist of only one Boson and one Fermion, and yield the short representations of (\ref{eq:bigalgebra1}--\ref{eq:bigalgebra2}) by tensor products.  The only physical input in this procedure is the identification of the eigenvalues of $M, P$ and $K$ --- while $H$ follows from the shortening condition --- in terms of the physical parameters of the theory. For the pure-RR case, this is  the amount of RR flux~$h$ (\textit{i.e.}, the string tension), and the particle's momentum $p$. As it turns out, for one particle states
\begin{equation}
M = \pm1 \,,\qquad
P = +\frac{ih}{2}\big(e^{ip}-1\big)\,,\qquad
K = -\frac{ih}{2}\big(e^{-ip}-1\big)\,,
\end{equation}
where the sign in $M$ identifies different irreducible representations (\textit{i.e.}, different families of particles).
As it turns out~\cite{Hoare:2013pma,Hoare:2013lja,Lloyd:2014bsa}, it is easy to accommodate the mixed-flux case in the above structure by leaving the algebra unchanged, and modifying the representation parameters by introducing the amount of NSNS flux $k$, which for $h=0$ becomes the quantised level of the WZW model:
\begin{equation}
\label{eq:chargesvalues}
M = \pm1 +\frac{k}{2\pi}p\,,\qquad
P = +\frac{ih}{2}\big(e^{ip}-1\big)\,,\qquad
K = -\frac{ih}{2}\big(e^{-ip}-1\big)\,.
\end{equation}
It is worth noting that the shortening condition~\eqref{eq:shortening} yields a \textit{chiral} dispersion relation for $h=0$. From now on and until further notice, let us assume that $h$ and $k$ are generic.

The S-matrix is the non-trivial invariant tensor on the two-particle representation
 of the algebra (\ref{eq:bigalgebra1}--\ref{eq:bigalgebra2}). Again, this can be worked out for the smaller algebra~\eqref{eq:smallalgebra}, as it was done in detail in ref.~\cite{Borsato:2012ud}, and then extended to the case of (\ref{eq:bigalgebra1}--\ref{eq:bigalgebra2}). As it turns out, for a non-trivial S~matrix to exist, it is necessary to endow the algebra~$\mathcal{A}$ of \eqref{eq:smallalgebra} with a non-trivial coproduct $\Delta: \mathcal{A}\to\mathcal{A}\otimes\mathcal{A}$, given by
\begin{equation}
\Delta(Q_{*}) =Q_{*}\otimes 1+ U^{+1}\otimes Q_{*}\,,\qquad\Delta(S_{*}) =S_{*}\otimes 1+ U^{-1}\otimes S_{*}\,,
\end{equation}
where $*$ stands everywhere either for $L$ or $R$, and $U^{\pm1}=e^{\pm ip/2}$. From this it follows that%
\footnote{Among other things, this guarantees that the eigenvalues of $P,K$ on multiparticle states with momenta $p_1, p_2, \dots p_n$depend only on the total momentum $p_1+\cdots +p_n$.}
\begin{equation}
\begin{gathered}
\Delta(P)=P\otimes 1+U^{+2}\otimes P\,,\qquad
\Delta(K)=K\otimes 1+U^{-2}\otimes K\,,\\
\Delta(H_{*})=H_{*}\otimes 1+1\otimes H_{*}\,.
\end{gathered}
\end{equation}

The above structure is sufficient to construct the S~matrix scattering a pair of irreducible short representation, which was done in~\cite{Borsato:2012ud,Hoare:2013lja,Lloyd:2014bsa}. As always, this up to an overall scalar factor, the dressing factor, see refs.~\cite{Janik:2006dc, Beisert:2006ez,Borsato:2013hoa, Borsato:2016xns}, for pair of irreducible representations. Here we have two such representations (one containing $z,y$ and their fermionic partners, one with $\bar{z}, \bar{y}$) and hence we expect \textit{four dressing factors}.
 Without delving too deep in that derivation, it is worth emphasising a consequence of the factorisation~\eqref{eq:algebrasplit}. The short representations of the larger algebra~(\ref{eq:bigalgebra1}--\ref{eq:bigalgebra2}) are obtained by taking tensor products of short representations of the smaller algebra~\eqref{eq:smallalgebra}, and as a consequence the full S~matrix has the form $S=\widetilde{S}\otimes\widetilde{S}$, where $\widetilde{S}$ is invariant under~\eqref{eq:smallalgebra}. As it turns out, $\widetilde{S}$ is a physical S~matrix (and not merely an auxiliary object) as it describes the scattering of worldsheet excitations of the $AdS_3\times S^3\times S^3\times S^1$ superstring~\cite{Borsato:2012ud}. Hence, all elements of $\widetilde{S}$ admit a perturbative expansion in large tension $h\gg1$. Depending on whether the S-matrix element is on the diagonal  we will have
\begin{equation}
\label{eq:largehsmatrix}
\widetilde{S}_{ij}^{ij}=1+O(h^{-1})\,,\qquad
\text{or}\quad
\widetilde{S}_{ij}^{kl}=O(h^{-1})\quad\text{if}\  
(k,l)\neq(i,j)\,.
\end{equation}
Given that each S-matrix element of $S$ is bilinear in the S-matrix elements of $\widetilde{S}$, it is possible to conclude that some of them must be of order~$O(h^{-2})$ solely due to the factorised structure~\eqref{eq:algebrasplit}. In fact, looking more closely at the form of the S~matrix as in ref.~\cite{Borsato:2013qpa} we find that it is a \textit{necessary condition} that the tree-level S~matrix is diagonal for the factorisation~\eqref{eq:algebrasplit} to hold. Hence, given the diagonal form of $T_{ij}^{kl}$ in eq.~\eqref{eq:Tisdiagonal}, we can see that also in the case at hand the symmetry algebra may factorise.

Let us now see how one may try to adjust the recipe above to accommodate the two parameter deformation with $\chi_\pm \neq 0$ (but $q=0$)~\cite{Hoare:2014oua}. First of all, as we are dealing with the $q=0$ case (no Wess-Zumino term) we set $k=0$ in eq.~\eqref{eq:chargesvalues}: this is what we want to deform. It is further assumed that each copy of $\su(1|1)$ is~\eqref{eq:bigalgebra1} is quantum-deformed in a standard way. Namely, we consider
\begin{equation}
\label{eq:bigalgebradef}
\{Q_{\sL}{}^i, S_{\sL}{}_{j} \} = \delta_{j}^{i}\, \frac{q_{\sL}{}^{+H_{\sL}}-q_{\sL}{}^{-H_{\sL}}}{q_{\sL}-q_{\sL}{}^{-1}},\qquad
\{Q_{\sR}{}_{i}, S_{\sR}{}^{j} \} = \delta_{i}^{j}\, \frac{q_{\sR}{}^{+H_{\sR}}-q_{\sR}{}^{-H_{\sR}}}{q_{\sR}-q_{\sR}{}^{-1}}\,.
\end{equation}
Like above it is equivalently effective and more convenient to restrict to the smaller algebra
\begin{equation}
\{Q_{\sL}, S_{\sL} \} = [H_{\sL}]_{q_{\sL}},\qquad
\{Q_{\sR}, S_{\sR} \} = [H_{\sR}]_{q_{\sR}}\,,
\end{equation}
where we introduced the notation $[X]_z = (z^X-z^{-X})(z-1/z)$.
The coproducts are also modified in the standard way,
\begin{equation}
\begin{aligned}
\Delta(Q_{*}) =&\,Q_{*}\otimes 1+ q_{*}^{H_{*}}U^{+1}\otimes Q_{*}\,,\qquad
&\Delta(q_{*}^{H_{*}})=&\,q_{*}^{H_{*}}\otimes q_{*}^{H_{*}}\\
\Delta(S_{*}) =&\,S_{*}\otimes q_{*}^{-H_{*}}+ U^{-1}\otimes S_{*}\,,\qquad&
\Delta(U)=&\,U\otimes U\,,
\end{aligned}
\end{equation}
where again $*$ stands either for $L$ or $R$ everywhere, from which it follows
\begin{equation}
\begin{gathered}
\Delta(P)=P\otimes 1+q_{\sL}^{H_{\sL}}q_{\sR}^{H_{\sR}}U^{+2}\otimes P\,,\qquad
\Delta(K)=K\otimes q_{\sL}^{-H_{\sL}}q_{\sR}^{-H_{\sR}}+U^{-2}\otimes K\,,\\
\Delta(H_{*})=H_{*}\otimes 1+1\otimes H_{*}\,.
\end{gathered}
\end{equation}
Finally, since the S~matrix must commute in particular with $K$ and $P$, it follows that these two generators must be co-commutative, which forces
\begin{equation}
P=\frac{\nu}{2}\big(1-q_{\sL}^{H_{\sL}}q_{\sR}^{H_{\sR}}U^2\big)\,,\qquad
K=\frac{\nu}{2}\big(q_{\sL}^{-H_{\sL}}q_{\sR}^{-H_{\sR}}-U^{-2}\big)\,,
\end{equation}
where $\nu$ is an undetermined proportionality constant. Finally, the representations in which we are interested will satisfy
\begin{equation}
\label{eq:shortening-qdef}
[H_{\sL}]_{q_{\sL}}\,[H_{\sR}]_{q_{\sR}}=P\,K\,.
\end{equation}

Under the above assumptions one can work out the all-loop S~matrix up to dressing factors~\cite{Hoare:2014oua}. In order to compare with the perturbative result, it is necessary to relate the parameters $q_{\sL}$, $q_{\sR}$ and $\nu$ to the perturbative parameters which we used: $\chi_{\pm}$, and the string tension~$h$, which we take to be large. The proposal of ref.~\cite{Hoare:2014oua} is
\begin{equation}
\nu=\frac{h}{\sqrt{1+\chi_+^2}}+O(h^0)\,,\qquad
q_{*}=e^{-\chi_{*}/h}+O(h^{-2})\,,\qquad
U=e^{i p/(2h)}+O(h^{-2})\,,
\end{equation}
while%
\footnote{%
The fact that the momentum scales as $1/h$ is typical of the near-BMN limit.}
 $\chi_{\pm}=\chi_{\sL}\pm\chi_{\sR}$. It is also assumed that $H_{\sL}$ and $H_{\sR}$ are given by their tree-level values. Plugging this in eq.~\eqref{eq:shortening-qdef}, we get at leading order
\begin{equation}
(\omega\mp\chi_+\chi_-)^2-p^2-m^2=0\,,
\end{equation}
with $m$ given by~\eqref{eq:quadraticH}, which precisely reproduces~\eqref{eq:omega} at $q=0$.
We have explicitely checked that applying the same expansion to the S-matrix elements of ref.~\cite{Hoare:2014oua} matches our perturbative computation.
Note that, since ref.~\cite{Hoare:2014oua} did not propose the four dressing factors of this S~matrix, we have checked $12=16-4$ independent ratios of S-matrix elements.

At this point, it is worth speculating on whether we can accommodate $q\neq0$ in this formalism. In the undeformed case, is is possible to account for the WZ term by switching on $k$ in eq.~\eqref{eq:chargesvalues}. In other words, the symmetry algebra is \textit{unchanged}, and only the representation coefficients are deformed.
It is easy to see that if we assume only the representation to be deformed like in \eqref{eq:chargesvalues}, already at the level of the dispersion relation we find that this does not reproduce the form of perturbative result. Some more general ans\"atze can accommodate the form of eq.~\eqref{eq:omega}. In particular, by allowing arbitrary linear shifts in momenta both in the identification of $H_{\sL}$  and $H_{\sL}$, as well as an arbitrary near-BMN scaling of $p$ in $U$ and of the constant term $m$ in $H_{\sL}-H_{\sR}$, we can reproduce the dispersion relation --- essentially by having sufficiently many free parameters --- yet not the three-parameter S~matrix.

\subsection{Chiral limits}
\label{sec:chiral}
By tuning the parameters $\chi_+,\chi_-$ and $q$ it is formally possible to make the dispersion relation~$\omega(p)$ \textit{chiral}, \textit{i.e.}\ to have that
\begin{equation}
\label{eq:chiralomega}
\frac{\partial }{\partial p} \omega(p) = \pm c\,,
\end{equation}
and similarly for $\bar{\omega}(p)$, for some constant $c\neq0$ --- the ``speed of light'' of these particles.%
\footnote{%
Our normalization of the dispersion relation~\eqref{eq:omega} sets $c=1$.}
In other words, we can make it so that the particles of the model move with constants speed either to the left or to the right --- at least at tree-level.
The interest in considering this limit is due to the fact that in such a massless relativistic theory%
\footnote{%
We use the term ``relativistic'' theory a little loosely, as the terms of the form \textit{e.g.} $\bar{Z}\acute{Z}$ and $\bar{Z}P_z$ in~\eqref{eq:quadraticH} break two-dimensional Poincar\'e symmetry. Such terms can emerge by coupling a Poincar\'e invariant theory to a background gauge field.}
the kinematics is quite restricted, and the S~matrix takes a very simple form. It is easy to see that we can satisfy~\eqref{eq:chiralomega} by $\lambda=\pm m$ in eq.~\eqref{eq:omega}. This gives three choices
\begin{equation}
\begin{aligned}
\label{eq:speciallimits}
&\text{case I:}\qquad &&\chi_+=i\,,\\
&\text{case II:}\qquad &&\chi_-=i\,,\\
&\text{case III:}\qquad &&q=i\,(\chi_+ + \chi_-)\,.
\end{aligned}
\end{equation}
All of these correspond to complex actions, but this will turn out not to be an issue, as we will see below. A few more related solutions follow from flipping the signs of $\chi_+$ and $\chi_-$.

For each of the choices in~\eqref{eq:speciallimits}, we want to look at the corresponding S~matrix. Given that the particles move at the speed of light, for the scattering to be well-defined in perturbation theory, we must choose the momenta of the two particles $p_1$ and $p_2$ so that
\begin{equation}
\frac{\partial \omega}{\partial p}\Big|_{p_1}=+c\,,
\qquad\frac{\partial \omega}{\partial p}\Big|_{p_2}=-c\,,
\end{equation}
where we assume for definiteness $c>0$, and similarly for $\bar{\omega}(p)$. If the resulting theory were chiral and relativistic, the scattering should only depend on the unique Mandelstam invariant~$s=-p_1p_2$. In our case, given that we allow for ``shifts'' in the dispersion similar to those that would emerge from a background gauge field, we should expect also linear and constant terms in $p_1$, $p_2$ --- like for instance the so-called frame factors~\cite{Arutyunov:2006yd}. It should be stressed that this tree-level analysis does not guarantee that the whole quantum theory will remain chiral, but rather should be seen as a way to single out a few particularly interesting families of theories in the \textit{mare magnum} of this three-parameter model.
We already encountered an example of this chiral scattering: the undeformed $AdS_3\times S^3$ Wess-Zumino-Witten model, whose S~matrix~\eqref{eq:WZWSmat} we ``rediscovered'' from a limit of the three-parameter case. Moreover, marginal current-current deformations of the WZW model~\cite{Gepner:1986wi,Chaudhuri:1988qb,Forste:2003km} should also exhibit a similar chiral structure.
As we briefly recalled around eq.~\eqref{eq:WZWSmat} and as analysed at length in refs.~\cite{Baggio:2018gct,Dei:2018mfl} (see also~\cite{Dei:2018jyj}), the all-loop S~matrix of such theories must have an even more constrained form than one might have imagined from kinematics. This very rigid structure is necessary to ensure that the spectrum constructed from the mirror thermodynamic Bethe ansatz~\cite{Dei:2018mfl} matches what one may derive from the Sugawara construction for the $\sl(2)\oplus\su(2)$ Ka\v{c}-Moody algebra. Referring the reader to~\cite{Dei:2018mfl,Dei:2018jyj} for details, we note that for the tree-level S~matrix the only allowed structure is that $T_{\alpha\beta}(p_1,p_2)$ should be a \textit{polynomial} in $p_1$ and $p_2$ of maximum degree~$1$, \textit{cf.}\ eq.~\eqref{eq:WZWSmat}. This is enormously restrictive, since even at tree-level we might have expected a meromorphic function of~$s$.%
\footnote{This form is also reminiscent of a $T\bar{T}$ deformation of a free theory~\cite{Cavaglia:2016oda,Baggio:2018gct}.}

\paragraph{Case I.}
Plugging $\chi_+=i$ into the metric and $B$-field (\ref{eq:defds}--\ref{eq:defA}) we find a complex action. To bring it to a real form, we change variables
\begin{equation}
\label{eq:complexchange}
r=\tilde{r}\,,\qquad\phi=\tilde{\phi}-i\frac{q}{2(1+\chi_-^2)}\log(\tilde{r}^2-1)\,,\qquad
\omega=\tilde{\omega}-\frac{q\chi_-}{2(1+\chi_-^2)}\log(\tilde{r}^2-1)\,,
\end{equation}
and similarly for the $t$ and $\psi$.%
\footnote{Recall that the $AdS_3$ and $S^3$ parts of the trice-deformed metric are related by analytic continuation, \textit{cf.}\ eq.~\eqref{eq:analyticcont}. In the remainder of this section, for brevity we will only discuss the change of coordinates on~$S^3$.}
Dropping the tildes from the new coordinates, we obtain the metric and $B$-field
\begin{equation}
\label{eq:caseIa}
\begin{aligned}
&\de s^2=
\frac{\de \rho^2}{(1+\rho^2)^2(1+\chi_-^2)}
+\frac{\rho^2\de\psi^2}{1+\chi_-^2-q^2\rho^2}-\frac{[1-\chi_-^2(1+\rho^2)]\de t^2}{1+\chi_-^2-q^2\rho^2}
+\frac{2i\chi_- \rho^2\de t\de\psi}{1+\chi_-^2-q^2\rho^2}
\\
&\quad+\frac{\de r^2}{(1-r^2)^2(1+\chi_-^2)}
+\frac{r^2\de\phi^2}{1+\chi_-^2+q^2r^2}+\frac{[1-\chi_-^2(1-r^2)]\de\omega^2}{1+\chi_-^2+q^2r^2}
+\frac{2i\chi_- r^2\de\omega\de\phi}{1+\chi_-^2+q^2r^2}\,,\\
&B=\frac{q\,\rho^2}{1+\chi_-^2-q^2\rho^2}\,\de t\wedge\de\psi\,,+\frac{q\,r^2}{1+\chi_-^2+q^2r^2}\,\de\omega\wedge\de\phi\,,
\end{aligned}
\end{equation}
where we assumed that $1+q^2+\chi_-^2>0$, which is true for the deformation parameters sufficiently close to zero. We see that~\eqref{eq:caseIa} is perfectly real for $i\chi_-\in\mathbb{R}$. At this point, to avoid any concern stemming from the complex change of coordinates~\eqref{eq:complexchange}, we can forget how we obtained~\eqref{eq:caseIa} and derive the tree level S~matrix from scratch from the real metric and $B$-field. We find that the dispersion relation is indeed chiral,
\begin{equation}
\label{eq:shift}
\omega(p)=\big|p-q\big|-i\chi_-\,,\qquad
\bar{\omega}(p)=\big|p+q\big|+i\chi_-\,,
\end{equation}
and that the bosonic tree-level S~matrix precisely coincides with what we would find by plugging~$\chi_+=i$ in appendix~\ref{app:Smatrix}. This background also solves the supergravity equations with all RR fluxes set to zero, with the dilaton
\begin{equation}
e^{2\Phi}=e^{2\Phi_0}\,\frac{(1+\rho^2)\,(1-r^2)}{(1+\chi_-^2-q^2\rho^2)\,(1+\chi_-^2+q^2r^2)}\,,
\end{equation}
where $\Phi_0$ is a constant dilaton.
In fact, it is possible to perform one more change of coordinates to see that~\eqref{eq:caseIa} is related to a current-current deformation of a $\sl(2)\oplus\su(2)$ WZW model~\cite{Gepner:1986wi,Chaudhuri:1988qb,Forste:2003km}. By further performing the shift
\begin{equation}
r=\tilde{r}\,,\qquad\phi=\tilde{\phi}-i\chi_-\tilde{\omega}\,,\qquad
\omega=\tilde{\omega}\,,
\end{equation}
and similarly for $AdS_3$, and by rescaling the coordinates, we can bring \eqref{eq:caseIa} to the metric and $B$-field of a current-current deformation of a WZW model, written in the coordinates of ref.~\cite{Forste:2003km}
\begin{equation}
\label{eq:currentcurrent}
\begin{aligned}
\de s^2 =&\,\zeta\,\Big[
\de \rho^2-\frac{\de t^2}{R^2-\coth^2r}+\frac{R^2\coth^2r\;\de \psi^2}{R^2-\coth^2r}\\
&\qquad\quad
 +\de r^2+\frac{\de \omega^2}{R^2+\cot^2r }+\frac{R^2\cot^2r\;\de \phi^2}{R^2+\cot^2r}\Big],\\
B =&\,\zeta\,\Big[ \frac{\coth^2r}{R^2-\coth^2r}\,\de t\wedge\de\psi+\frac{\cot^2r}{R^2+\cot^2r}\,\de\omega\wedge\de\phi\Big]\,,
\end{aligned}
\end{equation}
where $R^2=\zeta/q^2$ while $\zeta=-1-\chi_-^2$ is related to the level of the WZW model, and once again we dropped the tildes. Note that, even if the shift~\eqref{eq:shift} is real for $i\chi_-\in\mathbb{R}$, \textit{it does affect} the S~matrix, because it involves the lightcone coordinate~$\omega$ (and $t$ in the $AdS_3$ part).%
\footnote{%
See ref.~\cite{Sfondrini:2019smd} for a detailed discussion of the effects of such shift on the lightcone-gauge-fixed S~matrix.}
In this case%
\footnote{%
To carry out the perturbative computation of the S~matrix it is convenient to choose coordinates that allow for a perturbative weak-field expansion of the transverse fields. To this end, it is sufficient to set $r=\text{acot}(\tilde{r})$, $\omega=\tilde{\omega}$, $\phi=\tilde{\phi}$ and similarly for the $AdS_3$ coordinates.%
} the dispersion relation is modified to
\begin{equation}
\label{eq:simpleomegaI}
\omega(p) = \frac{1}{R^2}\,\big|\zeta\, p-1\big|\,\qquad
\bar{\omega}(p) = \frac{1}{R^2}\,\big|\zeta\, p+1\big|\,.
\end{equation}
Note that the scaling of the leading-order term in~$p$ is different from the three-parameter deformation --- the ``speed of light'' is $c=\zeta/R^2$ rather than $c=1$. This can be reabsorbed by rescaling the worldsheet coordinate~$\sigma$. The tree-level bosonic S~matrix in the gauge $\alpha=1/2$ takes the simple form
\begin{equation}
\label{eq:simpleTI}
T_{ij}(p_1, p_2) = \frac{1}{2R^2}\Big[
p_1 (J_\psi^{j}+J_\phi^{j})+
p_2 (J_\psi^{i}+J_\phi^{i})\Big]-\frac{R^2-1}{\zeta\,R^2} \big(J_\psi^{i}J_\psi^{j}-J_\phi^{i}J_\phi^{j}\big)\,.
\end{equation}
Once again, like in eq.~\eqref{eq:WZWSmat}, $J_{\psi}^{i,j}$ and $J_{\phi}^{i,j}$ are the eigenvalues of the $\u(1)$ charges of table~\ref{tab:charges} relative to the particle flavour $i$ or $j$.

\paragraph{Case II.}
We can relate this case to the previous one by recalling that --- up to flipping the sign of the $B$-field --- the three-parameter geometry is invariant under~\cite{Delduc:2018xug}
\begin{equation}
\chi_{\pm}\to\chi_{\mp}\,,\qquad
q\to-q\,,
\end{equation}
as long as we redefine the coordinates as it follows:
\begin{equation}
\rho\to i\sqrt{1+\rho^2}\,,\quad
t\to\psi\,,\quad
\psi\to t\,,\qquad
r\to\sqrt{1-r^2}\,,\quad
\omega\to\phi\,,\quad
\phi\to\omega\,.
\end{equation}
It thus follows that also this case yields a background related to a current-current deformation of a WZW model, up to a coordinate shift of the type~\eqref{eq:shift}.

\paragraph{Case III.}
In this case, we note that the S~matrix remains fairly involved for general values of $\chi_+$ and $\chi_{-}$. A simplification occurs when $\chi_+=i$ or $\chi_-=i$, which correspond to cases I and II, respectively. Another very interesting case, where a drastic simplification occurs, is when we set
\begin{equation}
\label{eq:newlimit}
q= i\,\chi_+\,,\qquad\chi_-=0\,,
\end{equation}
and similarly by exchanging the roles of $\chi_{+}$ and $\chi_-$ (see the discussion in case II).
Also in this case the action is complex and it is necessary to do a complex transformation
\begin{equation}
\label{eq:complexshiftII}
r=\tilde{r}\,,\qquad
\omega=\tilde{\omega}\,,\qquad
\phi=\tilde{\phi}+\frac{i}{2}\log\big(1+\chi_+^2\tilde{r}^2\big)\,,
\end{equation}
and similarly for~$AdS_3$. Again we take the resulting real action as our starting point, forgetting the formal manipulation~\eqref{eq:complexshiftII} and dropping the tildes. As it turns out, the result matches what we would have found by plugging~\eqref{eq:newlimit} in the three-parameter S~matrix. In particular, the dispersion is the same as in the underformed WZW model~\cite{Hoare:2013lja,Lloyd:2014bsa}
\begin{equation}
\label{eq:simpleomegaII}
\omega(p)= \big|p-1\big|\,,\qquad
\bar{\omega}(p) = \big|p+1\big|\,,
\end{equation}
and the tree level S~matrix can be expressed neatly in closed form. In the gauge $\alpha=1/2$ we find%
\footnote{%
It is simple to restore the gauge dependence in the tree-level S~matrix by adding a term $(\tfrac{1}{2}-\alpha)(p_1\omega_2-p_2\omega_1)$, see appendix~\ref{app:Smatrix} for details.}
\begin{equation}
\label{eq:simpleTII}
T_{ij}(p_1, p_2) = \frac{1}{2}\Big[
p_1 (J_\psi^{j}+J_\phi^{j})+
p_2 (J_\psi^{i}+J_\phi^{i})\Big]-2\chi_{+}^2 \big(J_\psi^{i}J_\psi^{j}-J_\phi^{i}J_\phi^{j}\big)\,.
\end{equation}
For $\chi_+=0$, we retrieve precisely the S-matrix of the undeformed WZW model in the $\alpha=1/2$ gauge~\cite{Hoare:2013pma,Baggio:2018gct}. In fact, for general $\chi_+$ we recognise that eqs.~(\ref{eq:simpleomegaII}--\ref{eq:simpleTII}) match eqs.~(\ref{eq:simpleomegaI}--\ref{eq:simpleTI}) up to rescaling~$\sigma$, $\tau$ and the string tension~$h$,\footnote{%
Recall that perturbatively for $h\gg1$ we have
$S_{ij}^{ij}=1+h^{-1}T_{ij}+O(h^{-2})$.
} and identifying
\begin{equation}
\label{eq:Rchirelation}
R^2=1+2\chi_+^2\,.
\end{equation}
 The corresponding background is%
\begin{equation}
\label{eq:newlimitII}
\begin{aligned}
\de s^2 =&\ 
\frac{\de \rho^2}{(1+\rho^2)(1-\chi_+^2\rho^2)}-
\frac{(1+\rho^2)\,\de t^2}{1+\chi_+^2\rho^4}+
\frac{\rho^2(1-\chi_+^2\rho^2)\,\de\psi^2}{1+\chi_+^2\rho^4}\\
&\qquad+\frac{\de r^2}{(1-r^2)(1+\chi_+^2r^2)}+\frac{(1-r^2)\,\de\omega^2}{1+\chi_+^2r^4}+\frac{r^2(1+\chi_+^2r^2)\,\de\phi^2}{1+\chi_+^2r^4}\,,\\
B=&\ 
\frac{\rho^2(1-\chi_+^2\rho^2)}{1+\chi_+^2\rho^4}\,\de t\wedge\de\psi+
\frac{r^2(1+\chi_+^2r^2)}{1+\chi_+^2r^4}\,\de\omega\wedge\de\phi\,.
\end{aligned}
\end{equation}
It is tempting to try to identify this background with the current-current deformation~\eqref{eq:currentcurrent}. To investigate this point it is convenient to change coordinates using Jacobi elliptic functions,
\begin{equation}
r=\text{sn}(\tilde{r}; -\chi_+^2)\,,\qquad
\omega=\tilde{\omega}\,,\qquad \phi=\tilde{\phi}\,,
\end{equation}
and similarly for $AdS_3$. Dropping the tildes we get, for the $S^3$ part,
\begin{equation}
\begin{aligned}
\de s^2_{(2)}=&\,\xi^2\, \de r^2+\frac{1-\text{sn}^2 (\xi\,r,-\chi_+^2)}{1+\chi_+^2\text{sn}^4 (\xi\,r,-\chi_+^2)}\,\de\omega^2+\frac{\text{sn}^2 (\xi\,r,-\chi_+^2)+\chi_+^2\text{sn}^4 (\xi\,r,-\chi_+^2)}{1+\chi_+^2\text{sn}^4 (\xi\,r,-\chi_+^2)}\,\de\phi^2\,,\\
B_{(2)}=&\,\frac{\text{sn}^2 (\xi\,r,-\chi_+^2)+\chi_+^2\text{sn}^4 (\xi\,r,-\chi_+^2)}{1+\chi_+^2\text{sn}^4 (\xi\,r,-\chi_+^2)}\,\de\omega\wedge\de \phi\,,
\end{aligned}
\end{equation}
and similarly for $AdS_3$. Note that for convenient we have introduced a factor of
\begin{equation}
\xi =\frac{2 K(-\chi_+^2)}{\pi}\,,
\end{equation}
where $K(z)$ denotes the complete elliptic integral of the first kind, so that the coordinate $r$ has a real period of length~$2\pi$.
Starting from this form and expanding the metric for small~$r\ll1$, in such a way as to retain only the terms needed for the quartic Hamiltonian, we find that indeed this background is diffeomorphic to~\eqref{eq:currentcurrent} in that limit, provided that we identify $\chi_+$ as in eq.~\eqref{eq:Rchirelation}. This explains the matching of the tree-level S~matrix, which in this light may well be a tree-level accident.
On the other hand, expanding in the deformation parameter~$\chi_+\ll1$ we find
\begin{equation}
\label{eq:expandedmetric}
\begin{aligned}
\de s^2_{(2)}=&\,\xi^2\, \de r^2+
\cos^2r\, \de\omega^2+
\sin^2r\, \de\phi^2
+\frac{\chi_+^2}{16}\,\sin 2r\,\sin 4r
\big(\de\omega^2-\de\phi^2\big)
+O(\chi_+^4),\\
B_{(2)}=&\,
\big(\sin^2r-\frac{\chi_+^2}{16}\,\sin 2r\,\sin 4r\big)
\de\omega\wedge\de \phi+O(\chi_+^4)\,,
\end{aligned}
\end{equation}
where $\xi=1-\chi_+^2/4+O(\chi_+^4)$. Even at small $\chi_+\ll1$ it is not clear how to match this background with a current-current deformation. Moreover, by direct inspection it appears that~\eqref{eq:expandedmetric} \textit{does not} solve the supergravity equations at order $O(\chi_+^2)$ (even in a generalised sense~\cite{Arutyunov:2015mqj,Wulff:2016tju}) in absence of RR fluxes.
It would be very interesting to complete the background by extracting the RR fluxes from~\cite{Delduc:2018xug}, which would require generalising the approach of ref.~\cite{Borsato:2016ose,Seibold:2019dvf}, and computing the scattering processes involving Fermions.
While it is possible that the simplicity of~$\eqref{eq:simpleTII}$ is an accident of the bosonic tree-level computation, typically the integrable structure together with unitarity heavily constrain the S~matrix, and it would be very interesting to see whether the scattering remains chiral and simple at higher orders. 

\section{Conclusions}
\label{sec:conclusions}

We have computed the tree-level bosonic S~matrix for the three-parameter deformation of ref.~\cite{Delduc:2018xug}.
We find that, when restricting to $q=0$, it is compatible with the all-loop form conjectured in ref.~\cite{Hoare:2014oua}.
It is not straightforward, however, to conjecture a three-parameter all-loop S~matrix from tweaking the quantum-deformed representations of ref.~\cite{Hoare:2014oua}. It would be interesting to work out the symmetries of the three-parameter action following ref.~\cite{Delduc:2014kha}, in such a way to have firmer a guiding principle for bootstrapping the S~matrix.
We have also studied limits where the tree-level dynamics becomes chiral, where we might expect the worldsheet theory to have a simple structure (like it happens in the WZW limit~\cite{Baggio:2018gct,Dei:2018mfl}). In this way, we encounter a background related by a shift of the isometric coordinates to the marginal current-current deformation of the WZW model, see \textit{e.g.}~\cite{Gepner:1986wi,Chaudhuri:1988qb,Forste:2003km}. Both the shifted and not-shifted S~matrices are therefore byproducts of our work. Another interesting limit is the geometry~\eqref{eq:newlimitII} for which the tree-level bsonic S~matrix is extremely simple, \textit{cf.}~eq.~\eqref{eq:simpleTII}. In fact, at this order the S~matrix coincides with that of the current-current deformation up to some appropriate identifications. However it appears that, unlike the current-current deformation, the background~\eqref{eq:newlimitII} is not a supergravity solution (or a generalised supergravity solution~\cite{Arutyunov:2015mqj,Wulff:2016tju}) in absence of RR fluxes already at leading order in the deformation parameter. It is possible that the RR fluxes would drastically alter the nice structure~\eqref{eq:simpleTII} when considering Fermion scattering and beyond tree level. It would be interesting to extract the fluxes from the general three-parameter deformation by generalizing the techniques of ref.~\cite{Borsato:2016ose,Seibold:2019dvf}, firstly to verify whether they provide a supergravity background either in the ordinary or in the generalised~\cite{Arutyunov:2015mqj,Wulff:2016tju} sense, and secondly to work out the full S~matrix in the limit~\eqref{eq:newlimitII}.
We hope to return to some of these questions in the future.

\acknowledgments

We would like to thank Riccardo Borsato, Ben Hoare and Stijn J.~van Tongeren for insightful related discussions and for comments on a draft of this manuscript.
AS's work is funded by ETH Career Seed Grant No. SEED-2319-1 and by the Swiss National Science Foundation Spark grant  n.~190657.
FS  and AS are supported by the Swiss National Science Foundation through the NCCR SwissMAP.

\newpage
\appendix
\section{Quartic Hamiltonian}
\label{app:quarticH}
Below we write the quartic Hamiltonian $H^{(4)}$ for $\alpha=1/2$. We shall see in appendix~\ref{app:Smatrix}
 that the other values of $\alpha$ can be easily accounted for in the S~matrix.
\begin{equation}
\begin{aligned}
&H^{(4)} = \  \frac{1}{2} \Bigl\{m^2 \Bigl( -2 P_z \bar{P}_z Y \bar{Y} + 2 P_y \bar{P}_y Z \bar{Z} + 2 \acute{Y} \acute{\bar{Y}} Z \bar{Z} - 2 Y \bar{Y} \acute{Z} \acute{\bar{Z}}\\
&+\chi_{-} (q - i \chi_{+})\Bigl(+ \bar{P}_z Y \bar{Y} Z -\bar{P}_y Y Z \bar{Z}\Bigr) + \chi_{-} (q + i \chi_{+})\Bigl( P_z Y \bar{Y} \bar{Z} - P_y \bar{Y} Z \bar{Z} \Bigr)\Bigr)\\
&+ \lambda\Bigl(- i \bar{P}_y^2 Y \acute{Y} - i P_z \bar{P}_z \acute{Y} \bar{Y} +i P_z \bar{P}_z Y \acute{\bar{Y}}+ i P_y^2 \bar{Y} \acute{\bar{Y}} - i \acute{Y}^2 \bar{Y} \acute{\bar{Y}}  + i Y \acute{Y} \acute{\bar{Y}}^2\\
& + i \bar{P}_y \bar{P}_z \acute{Y} Z + i P_y \bar{P}_z \acute{\bar{Y}} Z  - i \bar{P}_y \bar{P}_z Y \acute{Z} + i P_y \bar{P}_z \bar{Y} \acute{Z}+i \bar{P}_z^2 Z \acute{Z} - i \bar{P}_y P_z \acute{Y} \bar{Z} \\
& -i P_y P_z \acute{\bar{Y}} \bar{Z}  + i P_y \bar{P}_y \acute{Z} \bar{Z} + i \acute{Y} \acute{\bar{Y}} \acute{Z} \bar{Z} - i \bar{P}_y P_z Y \acute{\bar{Z}} + i P_y P_z \bar{Y} \acute{\bar{Z}} - i P_y \bar{P}_y Z \acute{\bar{Z}}\\
&-i \acute{Y} \acute{\bar{Y}} Z \acute{\bar{Z}} - i \acute{Y} \bar{Y} \acute{Z} \acute{\bar{Z}} + i Y \acute{\bar{Y}} \acute{Z} \acute{\bar{Z}} - i P_z^2 \bar{Z} \acute{\bar{Z}} + i \acute{Z}^2 \bar{Z} \acute{\bar{Z}}- i Z \acute{Z} \acute{\bar{Z}}^2 \\
&+(m^2 - 2 i q \chi_{-}^2 \chi_{+})\Bigl( - i Y \acute{\bar{Y}} Z \bar{Z} + i Y \bar{Y} Z \acute{\bar{Z}} \Bigr) - \chi_{-} (i q - \chi_{+})\Bigl(2 \bar{P}_z Z \acute{Z} \bar{Z} - 2 \bar{P}_y Y \acute{Y} \bar{Y} \Bigr) \\
&+ (m^2 + 2 i q \chi_{-}^2 \chi_{+})\Bigl(+ i \acute{Y} \bar{Y} Z \bar{Z}  - i Y \bar{Y} \acute{Z} \bar{Z}\Bigr) +  \chi_{-} (iq + \chi_{+}) \Bigl( 2 P_z Z \bar{Z} \acute{\bar{Z}}-2  P_y Y \bar{Y} \acute{\bar{Y}}\Bigr)\\
&+(\chi_{-} \chi_{+}) \Bigl( - 2 \bar{P}_z Y \bar{Y} \acute{Z} + 2 \bar{P}_y \acute{Y} Z \bar{Z} + 2 P_y \acute{\bar{Y}} Z \bar{Z} - 2 P_z Y \bar{Y} \acute{\bar{Z}} \Bigr) \Bigr) \\
&+\chi_{-} (q - i \chi_{+}) \Bigl(-P_y^2 \bar{P}_y \bar{Y} - P_y P_z \bar{P}_z \bar{Y} + \bar{P}_y \acute{Y}^2 \bar{Y}  + \bar{P}_z \acute{Y} \bar{Y} \acute{Z}+ P_y \bar{P}_y P_z \bar{Z}\\
&+ P_z^2 \bar{P}_z \bar{Z} + P_z \acute{Y} \acute{\bar{Y}} \bar{Z}  - \bar{P}_y \acute{Y} \acute{Z} \bar{Z}-P_y \acute{\bar{Y}} \acute{Z} \bar{Z}  - \bar{P}_z \acute{Z}^2 \bar{Z} + P_z \acute{Y} \bar{Y} \acute{\bar{Z}} - P_y \bar{Y} \acute{Z} \acute{\bar{Z}} \Bigr)\\
&+  4 (1 + \chi_{-}^2 (2 + \chi_{+}^2))\Bigl(P_z \bar{P}_z Z \bar{Z} -P_y \bar{P}_y Y \bar{Y} \Bigr) + \chi_{-} (q + i \chi_{+}) \Bigl( - P_y \bar{P}_y^2 Y  - \bar{P}_y P_z \bar{P}_z Y\\
&+ P_y Y \acute{\bar{Y}}^2 + P_y \bar{P}_y \bar{P}_z Z + P_z \bar{P}_z^2 Z + \bar{P}_z \acute{Y} \acute{\bar{Y}} Z + \bar{P}_z Y \acute{\bar{Y}} \acute{Z} + P_z Y \acute{\bar{Y}} \acute{\bar{Z}} - \bar{P}_y \acute{Y} Z \acute{\bar{Z}} \\
&- P_y \acute{\bar{Y}} Z \acute{\bar{Z}}- \bar{P}_y Y \acute{Z} \acute{\bar{Z}} - P_z Z \acute{\bar{Z}}^2\Bigr) + (q^2 - \chi_{-}^2 + \chi_{+}^2)\Bigl( - 4 Y \acute{Y} \bar{Y} \acute{\bar{Y}} +4 Z \acute{Z} \bar{Z} \acute{\bar{Z}}  \Bigr) \\
&+ q \chi_{-}^2 \chi_{+}\Bigl( - 2 i P_y \bar{P}_z \bar{Y} Z  + 2 i \bar{P}_y P_z Y \bar{Z} - 2 i Y \acute{\bar{Y}} \acute{Z} \bar{Z} + 2 i \acute{Y} \bar{Y} Z \acute{\bar{Z}} \Bigr)  \\
&  +(q  \lambda \chi_{-} ) \Bigl(- 2 i \bar{P}_z Y \acute{\bar{Y}} Z + 2 i P_z \acute{Y} \bar{Y} \bar{Z} - 2 i P_y \bar{Y} \acute{Z} \bar{Z} + 2 i \bar{P}_y Y Z \acute{\bar{Z}}\Bigr)  \\
&+  (1 + \chi_{-}^2 + (q - i \chi_{+}) (q - i (1 + \chi_{-}^2) \chi_{+}))\Bigl(\acute{Y}^2 \bar{Y}^2 +P_z^2 \bar{Z}^2 -P_y^2 \bar{Y}^2- \acute{Z}^2 \bar{Z}^2\Bigr) \\
& +(1 + \chi_{-}^2 + (q + i \chi_{+}) (q + i (1 + \chi_{-}^2) \chi_{+}))\Bigl(- \bar{P}_y^2 Y^2  + Y^2 \acute{\bar{Y}}^2+ \bar{P}_z^2 Z^2 - Z^2 \acute{\bar{Z}}^2 \Bigr)  \\
& +(q^4 + 2 q^2 (1 + \chi_{+}^2) + (1 + \chi_{-}^2)^2 (-1 + \chi_{+}^4)) \Bigl(-2 Y^2 \bar{Y}^2  + 2 Z^2 \bar{Z}^2\Bigr)\\
&+ \chi_{-} ( q (q^{2} +  4 i q \chi_{+} + 1 + 5 \chi_{+}^2 + \chi_{-}^2 (-3 + \chi_{+}^2)) - 3 i\chi_{+}m^{2})\Bigl(P_z Z \bar{Z}^2 - P_y Y \bar{Y}^2  \Bigr)\\
& + \chi_{-}(q (q^{2} - 4 i q \chi_{+} +  1 + 5 \chi_{+}^2 + \chi_{-}^2 (-3 + \chi_{+}^2)) + 3 i\chi_{+}m^{2})\Bigl(\bar{P}_z Z^2 \bar{Z}  - \bar{P}_y Y^2 \bar{Y}  \Bigr)\\
& +a q (3 q^4 - 2 i q^3 \chi_{-}^2 \chi_{+} - 2 i q \chi_{-}^2 \chi_{+} (2 + \chi_{-}^2 + \chi_{+}^2) + q^2 (9 + 6 \chi_{+}^2 - \chi_{-}^2 (-2 + \chi_{+}^2))\\ 
&- (1 + \chi_{+}^2) (2 + \chi_{-}^4 - 3 \chi_{+}^2 + \chi_{-}^2 (7 + \chi_{+}^2)))\Bigl( + i Y^2 \bar{Y} \acute{\bar{Y}}- i Z^2 \bar{Z} \acute{\bar{Z}} \Bigr)\\
& + a q (3 q^4 + 2 i q^3 \chi_{-}^2 \chi_{+} + 
2 i q \chi_{-}^2 \chi_{+} (2 + \chi_{-}^2 + \chi_{+}^2) + q^2 (9 + 6 \chi_{+}^2 - \chi_{-}^2 (-2 + \chi_{+}^2))\\
& - (1 + \chi_{+}^2) (2 + \chi_{-}^4 - 3 \chi_{+}^2 + \chi_{-}^2 (7 + \chi_{+}^2)))\Bigl( - i Y \acute{Y} \bar{Y}^2 + i Z \acute{Z} \bar{Z}^2  \Bigr)\Bigr\}.
\end{aligned} 
\end{equation}

\section{Tree-level S~matrix}
\label{app:Smatrix}
The S~matrix allows for a perturbative expansion in the string tension~$h\gg1$, of the form
\begin{equation}
S_{ij}^{kl}= \delta_{i}^k\,\delta_j^l+\frac{i}{h}T_{ij}^{kl}+O(h^{-2})\,.
\end{equation}
Moreover, as discussed around~\eqref{eq:Tisdiagonal}, the tree-level S~matrix is diagonal, $T_{ij}^{kl}=\delta_{i}^k\delta_j^l\, T_{kl}$. The explicit for of the S-matrix elements is, at $\alpha=1/2$,
\begin{equation}
\begin{split}
    T_{ZY} = \frac{1}{2 D} & \Big[ m^{2} (p_{2}^{2} - p_{1}^{2}) +  \lambda (p_{1} - p_{2}) \Big( m^{2} + \chi_{-}^{2}\chi_{+}^{2} + p_{1}p_{2} - \omega_{1}\omega_{2}  \Big)\\
    & + \chi_{-} \chi_{+} \Big(p_{2} + p_{1} - 2 \lambda \Big) \Big( p_{1}(\omega_{2} + \chi_{-} \chi_{+}) - p_{2}(\omega_{1} + \chi_{-} \chi_{+}) \Big) 
    \Big],
\end{split}
\end{equation}
\begin{equation}
\begin{split}
    T_{Z\bar{Y}} =  \frac{1}{2 D} & \Big[m^{2} (p_{2}^{2} - p_{1}^{2}) + \lambda (p_{1} + p_{2}) \Big( m^{2} + \chi_{-}^{2}\chi_{+}^{2} - p_{1}p_{2} + \omega_{1}\bar{\omega}_{2}  \Big)\\  
    & + \chi_{-} \chi_{+} \Big(p_{1} - p_{2} - 2 \lambda \Big) \Big(  - p_{1}(\bar{\omega}_{2} - \chi_{-} \chi_{+}) + p_{2}(\omega_{1} + \chi_{-} \chi_{+}) \Big) 
    \Big],
\end{split}
\end{equation}
\begin{equation}
\begin{split}
    T_{\bar{Z}Y} =  \frac{1}{2 D} & \Big[ m^{2}(p_{2}^{2} - p_{1}^{2}) - \lambda (p_{1} + p_{2}) \Big( m^{2} + \chi_{-}^{2}\chi_{+}^{2}  - p_{1}p_{2} +\bar{\omega}_{1} \omega_{2}  \Big) \\
    & + \chi_{-} \chi_{+} \Big( p_{2} - p_{1} - 2 \lambda \Big) \Big( - p_{1}(\omega_{2} + \chi_{-} \chi_{+}) + p_{2}(\bar{\omega}_{1} - \chi_{-} \chi_{+}) \Big) 
    \Big],
\end{split}
\end{equation}
\begin{equation}
\begin{split}
    T_{\bar{Z}\bar{Y}} = \frac{1}{2 D} & \Big[ m^{2} (p_{2}^{2} - p_{1}^{2}) - \lambda  (p_{1} - p_{2}) \Big( m^{2} + \chi_{-}^{2}\chi_{+}^{2} + p_{1}p_{2} - \bar{\omega}_{1}\bar{\omega}_{2}  \Big) \\
    & + \chi_{-} \chi_{+} \Big(p_{1} + p_{2} + 2 \lambda \Big) \Big( - p_{1}(\bar{\omega}_{2} - \chi_{-} \chi_{+}) + p_{2}(\bar{\omega}_{1} - \chi_{-} \chi_{+}) \Big) 
    \Big],
\end{split}
\end{equation}
%
\begin{equation}
\begin{split}
    T_{ZZ} =  & \frac{1}{2 D} \Big[  (p_{1} + p_{2})^{2} + (q^{2} + \chi_{-}^{2} + \chi_{+}^{2})(\omega_{1} + \omega_{2})^{2} - 4 q^{2} (\omega_{1} \omega_{2} - p_{1} p_{2} -1)\\
    & - \chi_{-} \chi_{+} ( 4(\omega_{1} + \omega_{2}) + 4 \chi_{-}\chi_{+}(1 - \omega_{1} \omega_{2}) + (p_{2}\omega_{1} - p_{1}\omega_{2})(p_{2} - p_{1}) )\\
    & - \lambda \Big( p_{1} (p_{2}^{2} - \omega_{2} (\omega_{1} + 2 \chi_{-} \chi_{+}) + p_{2} (p_{1}^{2} - \omega_{1} (\omega_{2} + 2 \chi_{-} \chi_{+})))    \Big)\\
    & - aq(p_{1} + p_{2}) C
    \Big],
\end{split}
\end{equation}
\begin{equation}
\begin{split}
    T_{\bar{Z}\bar{Z}} = & \frac{1}{2 D} \Big[  (p_{1} + p_{2})^{2} + (q^{2} + \chi_{-}^{2} + \chi_{+}^{2})(\bar{\omega}_{1} + \bar{\omega}_{2})^{2} - 4 q^{2} (\bar{\omega}_{1} \bar{\omega}_{2} - p_{1} p_{2} -1)\\
    & + \chi_{-} \chi_{+} ( 4(\bar{\omega}_{1} + \bar{\omega}_{2}) - 4 \chi_{-}\chi_{+}(1 - \bar{\omega}_{1} \bar{\omega}_{2}) + (p_{2} \bar{\omega}_{1} - p_{1} \bar{\omega}_{2})(p_{2} - p_{1}) )\\
    &+ \lambda \Big(    p_{1} (p_{2}^{2} - \bar{\omega}_{2} (\bar{\omega}_{1} - 2 \chi_{-} \chi_{+}) ) + p_{2} (p_{1}^{2} - \bar{\omega}_{1}(\bar{\omega}_{2} - 2 \chi_{-} \chi_{+}))    \Big)\\
    & + aq(p_{1} + p_{2}) C
    \Big],
\end{split}
\end{equation}
\begin{equation}
\begin{split}
    T_{\bar{Z}Z} = & \frac{1}{2 D} \Big[  - (p_{1} - p_{2})^{2} - (q^{2} + \chi_{-}^{2} + \chi_{+}^{2})(\bar{\omega}_{1} - \omega_{2})^{2}  - 4 q^{2} (\bar{\omega}_{1} \omega_{2} - p_{1} p_{2} + 1)\\
   & + \lambda \Big( -p_{1}(p_{2}^{2} + \omega_{2} (\bar{\omega}_{1} - 2 \chi_{-} \chi_{+})) + p_{2}(p_{1}^{2} + \bar{\omega}_{1} (\omega_{2} + 2 \chi_{-} \chi_{+})) \Big) \\ 
   & + \chi_{-} \chi_{+} \Big( \omega_{2} (4 + p_{1}^{2} + p_{1} p_{2})  -  \bar{\omega}_{1} (4 + p_{2}^{2} + p_{1} p_{2}) + 4 \chi_{-} \chi_{+} (1 + \omega_{2} \bar{\omega}_{1}) \Big)  \\
    & + aq( - p_{1} + p_{2}) C
    \Big],
\end{split}
\end{equation}
where we introduced the short-hand,
\begin{equation}
    D = ( (p_{2} \pm \lambda )(\Omega_{1} \pm \chi_{-} \chi_{+} ) - (p_{1} \pm \lambda )  ( \Omega_{2} \pm \chi_{-} \chi_{+} ) ),
\end{equation}
where the sign of the shift $ \pm \lambda $ is positive for $Z$ and $Y$ and negative for $\bar{Z}$ and $\bar{Y}$ and $\omega_j$ is either $\omega(p_j)$ or $\bar{\omega}(p_j)$ depending on the particle's flavour. Similarly, we introduce the constant $C$ 
\begin{equation}
    C = q^{2} (7 + q^{2} + 2 (\chi_{-}^{2} + \chi_{+}^{2})) + (2 - \chi_{-}^{2} - \chi_{+}^{2} + \chi_{-}^{4} + \chi_{+}^{4} - 6 \chi_{-}^{2} \chi_{+}^{2})\,.
\end{equation}
The remaining S-matrix elements can be computed either by braising unitarity, which at tree level takes the form
\begin{equation}
T_{ij}(p_1,p_2) = - T_{ji}(p_2,p_1)\,,
\end{equation}
or by observing the relation between $AdS_3$ and $S^3$ fields, 
\begin{equation}
    S_{YY} = - S_{\bar{Z}\bar{Z}}\,,\qquad
    S_{\bar{Y}\bar{Y}} = - S_{ZZ}\,,\qquad
    S_{\bar{Y}Y} = - S_{Z\bar{Z}}\,,
\end{equation}
which is consequence of eq.~\eqref{eq:analyticcont}.

As it is well known from the literature on the uniform lightcone gauge~\cite{Arutyunov:2004yx,Arutyunov:2005hd,Arutyunov:2006gs} and as it has been discussed more recently in the context of $T\bar{T}$ deformations~\cite{Sfondrini:2019smd}, changing $\alpha$ only results in a multiplicative prefactor in the all-loop S~matrix, \textit{i.e.}, $S_{ij}^{kl}(p_1,p_2)\to e^{i(\alpha-1/2)\Theta(p_1,p_2)}\,S_{ij}^{kl}(p_1,p_2)$, with $\Theta(p_1,p_2) = p_2H(p_1)-p_1H(p_2)$ where $H(p)$ is the all-loop dispersion relation. Therefore, for the tree-level S~matrix, we have simply that at general $\alpha$
\begin{equation}
T_{ij}(p_1,p_2) = T_{ij}(p_1,p_2)\Big|_{\alpha=1/2}+(\alpha-\tfrac{1}{2})\Big(p_2\Omega(p_1)-p_1\Omega(p_2)\Big)\,.
\end{equation}

\bibliographystyle{JHEP}
\bibliography{refs}

\end{document}